\documentclass[12 pt]{osumkt-article}
\usepackage[utf8]{inputenc}
\usepackage[english]{babel}
\usepackage[margin=1in]{geometry}

\usepackage{amsmath}
\usepackage{graphicx}
\usepackage{booktabs}

\PassOptionsToPackage{hyphens}{url}\usepackage[colorlinks=true, allcolors=blue]{hyperref} 

\usepackage{tikzit}
\usepackage{comment}
\usepackage{booktabs} 
\usepackage{threeparttable} 
\usepackage{lscape} 

\usepackage{adjustbox} 

\usepackage{setspace}
\usepackage{natbib}
 \bibpunct[, ]{(}{)}{,}{a}{}{,}%
 \def\bibfont{\small}%

\usepackage{array,ragged2e}
\usepackage{caption} 
\newcolumntype{P}[1]{>{\RaggedRight\arraybackslash}p{#1}}

\usepackage{booktabs} 

\title{Unraveling Consumer Purchase Journey Using Neural Network Models}
\author{Victor Churchill\\Trinity College\\victor.churchill@trincoll.edu\\ \\H. Alice Li\\The Ohio State University\\li.815@osu.edu\\ \\Dongbin Xiu\\The Ohio State University\\xiu.16@osu.edu\\}

\begin{document}

\maketitle
\begin{abstract}

    \vspace{1em}
\begin{flushleft}
This study utilizes an ensemble of feedforward neural network models to analyze large-volume and high-dimensional consumer touchpoints and their impact on purchase decisions. When applied to a proprietary dataset of consumer touchpoints and purchases from a global software service provider, the proposed approach demonstrates better predictive accuracy than both traditional models, such as logistic regression, naive Bayes, and $k$-nearest neighbors, as well as ensemble tree-based classifiers, such as bagging, random forest, AdaBoost, and gradient boosting. By calculating the Shapley values within this network, we provide nuanced insights into touchpoint effectiveness, as we not only assess the marginal impact of diverse touchpoint types but also offer a granular view of the impact distribution within a touchpoint type. Additionally, our model shows excellent adaptability and resilience with limited data resources. When the historical data is reduced from 40 to 1 month, our model shows only a modest 19\% decrease in accuracy. This modeling framework can enable managers to more accurately and comprehensively evaluate consumer touchpoints, thereby enhancing the effectiveness and efficiency of their marketing campaigns.
\vspace{2em}


\end{flushleft}
\end{abstract}

\section{Introduction}\label{sec:intro}

Understanding consumers’ purchase journey has been a cornerstone of marketing research for over a century \citep{strong1925psychology}, evolving from early classic models like AIDA \citep{howard1969theory} to leveraging technological advancements for gaining deeper insights from more granular consumer data. The rise of online shopping has introduced various types of digital touchpoints, reflecting consumers’ underlying preferences and needs at different stages of their journey \citep{lemon2016understanding}. By analyzing these touchpoints across multiple marketing channels, firms can significantly improve their understanding of consumer behavior.

Accurately modeling and understanding the customer journey presents significant challenges. The analysis of online touchpoint data is particularly difficult due to its high dimensionality and large volume. This is further complicated by the increasingly diverse options for multi-channel shopping journeys. Firms struggle to extract meaningful patterns and insights from the vast amounts of touchpoints, and marketing managers are grappling with how to utilize touchpoint data in predicting consumer purchases or determining the impact of a single touchpoint on a customer’s journey towards purchase.

Additionally, traditional statistical models often require subjective categorization of touchpoints to reduce the data dimension. For example, touchpoints might be classified into Paid, Owned, and Earned media categories \citep{stephen2012effects} or distinguished by whether they are initiated by the firm or the consumer \citep{wiesel2011practice}. Moreover, stages like those outlined in the AIDA model are often predefined by researchers or managers \citep[e.g.,][]{hoban2015effects}. These predefined rules are based on subjective judgment rather than more flexible methodologies that adapt and evolve based on data patterns without being constrained by predetermined assumptions.

Our work enriches the literature by showcasing the application of a feedforward neural network framework on a unique proprietary dataset of consumer touchpoints and purchases. While existing research has shown neural networks' efficacy in processing unstructured data such as text and images \citep{zhang2022makes,tirunillai2014mining}, our study extends the use of neural networks to analyzing customer touchpoint data. In particular, our research demonstrates how neural networks can analyze complex patterns within large volumes of touchpoint data, surpassing state-of-the-art models in predicting customer conversions. It also enables firms to measure the relative importance of each touchpoint types, and further quantify the impact of each touchpoint within a touchpoint type. 

Leveraging the sophisticated capabilities of neural networks, our model allows for automated and objective categorization of touchpoints by the algorithm itself, moving beyond the confines of the predefined stages or touchpoint classifications. It adeptly determines the optimal network structure, including the number of latent layers and the nodes on each latent layer, to intricately map touchpoint interactions of the consumer's purchase journey, which manual methods might miss. Interestingly, for the context we studied, a simpler, shallower network structure emerged as most effective, highlighting our framework's adaptability to the data's specific characteristics.

Our model demonstrates outstanding performance in managing high-dimensional data, delivering superior predictive accuracy. We showcase the model’s effectiveness and precision by applying it to a dataset from a global software company. Our approach well handles the complexity in the data, showing both the highest capability to distinguish between buyers and non-buyers and the highest balanced accuracy in comparison with various established binary classifiers like logistic regression, naive Bayes, k-nearest neighbors, and advanced ensemble tree-based classifiers, including bagging, random forest, AdaBoost, and gradient boosting. In particular, our model achieves a notable Area Under the ROC\footnote{The receiver operator characteristic (ROC) curve plots the true positive rate against the false positive rate at different thresholds $\tau$.} Curve (AUROC) of 83.8\%, and 78.2\% accuracy in predicting buyers and 77.0\% in identifying non-buyers. 

We leverage Shapley values \citep{shapley1953value} to assess the impact of 31 touchpoint types in our context on consumer purchases, providing the focal firm with critical insights into the importance of each touchpoint type as done in previous attribution models \citep[e.g.,][]{li2014attributing, kireyev2016display}. Meanwhile, we also detail the distribution of touchpoints within a type, enabling a comprehensive view of their influence on purchasing decisions.

Handling high-dimensional touchpoint data is costly, yet our study demonstrates the model's robust performance with as little as one month of historical data. This adaptability is crucial for firms with limited resources, striking a balance between ease of data management and predictive accuracy. With a one-month lookback period, the model achieves an AUROC of 67.5\%, a modest decrease given the significantly reduced data window. This efficiency allows firms to make quick, informed decisions, enhancing strategic agility while minimizing data management efforts.

\section{Literature Review}\label{sec:lit}

Previous studies have examined the consumer purchase journey through the difference-in-differences model\citep{ghose2016toward}, logistic regression \citep{li2014attributing}, vector error correction model \citep{kireyev2016display}, hidden Markov model \citep{montgomery2004modeling}, point process \citep{xu2014path}, among other models. These conventional approaches, however, struggle with the complexity that comes from an increased number of channels and touchpoints, leading to a large parameter space and difficult estimations.

Machine learning models are increasingly utilized to analyze consumer journeys \citep{ma2020machine}. For example, \cite{trusov2016crumbs} use topic models to uncover latent themes from consumer clickstream data, and \cite{li2020charting} combine topic models and hidden Markov models to analyze search keywords used by consumers on their purchase journey. Additionally, \cite{liu2021content} employ Poisson factorization in interpreting the volumes of search queries and click-through rates.

Our research utilizes a feedforward neural network \citep{hastie2009elements} to delve into consumer touchpoints and their impact on purchase decisions. The marketing literature increasingly recognizes the power of neural networks, which allows for deep insights into various aspects, such as consumer reviews \citep{alantari2022empirical}, price targeting \citep{smith2023optimal}, ad recommendations \citep{aramayo2023multiarmed}, and more. This study aims to showcase the neural network's proficiency in navigating high-dimensional touchpoint data and handling complex nonlinear patterns, thereby greatly enhancing the predictive accuracy of purchases and contributing to our understanding of consumer purchase journeys.

Our research utilizes Shapley values within neural networks, enhancing the understanding of each touchpoint type's contribution to consumer purchases. We extend extant attribution models \citep{li2014attributing, kireyev2016display} by providing more nuanced insights. In addition to quantifying each touchpoint type's marginal impact on consumer purchases by extant attribution models, we can also delve into the distributions of the impact within each touchpoint type. As a result, the comprehensive assessment of touchpoint effectiveness on purchasing decisions and granular impact within a type of touchpoint offers managers a valuable tool to understand the potential over-utilization within certain touchpoint categories, guiding more accurate and targeted marketing decisions.

\section{Data}

Our data comprises a random sample of $20,556$ unique users obtained from a U.S.-based multinational software service provider over a period of 40 months, from June 2018 to September 2021. This focal firm offers a diverse range of over 20 software applications designed to support the creation and publication of content across various formats, including text, graphics, animation, video, and audio. Consumers are provided with annual subscription options to access these software services.

For each user in our dataset, we track their purchase activity along with the touchpoints that precede such purchases. Out of the total user base, 2,425 individuals have engaged in at least one purchase, constituting 11.8\% of all unique users in our dataset. We cease data collection following a user’s first purchase. Notably, our dataset has a pronounced class imbalance within the binary purchase decision. More specifically, our dataset contains a considerably greater number of non-buyers than buyers. 

Regarding the touchpoints, we have observed 31 distinct types of marketing touchpoints across all users. These include a wide range of activities from impressions and clicks on display ads, to engagements with email marketing through exposures and click-throughs, to interactions with both organic and paid search ads. A comprehensive description of these marketing touchpoints is detailed in Table \ref{table:touchpoints}. We have noticed a significant variability in the frequency of these touchpoints, with occurrences ranging from roughly 80 to 7 million over all users throughout the 40-month period. This extensive variability in touchpoint frequency typically poses challenges to traditional models. Researchers often attempt to mitigate these challenges by aggregating the data into broader categories, albeit at the expense of granularity. In our approach, we demonstrate how our model remains robust in the face of data imbalance.

\begin{table}
\centering
\caption{Touchpoints Types and Counts}
\begin{tabular}{l c c}
    \hline
    Touchpoint Type & Overall Count \\
    \hline
    
    Paid social impression & 30402 \\
    Paid social click & 78 \\
    Owned social click & 5167\\
    Earned social click & 12495\\    
    Paid affiliate click & 2178 \\
    Display impression type 1 & 3278563\\
    Display impression type 2 & 194759\\
    Display impression type 3 & 6995656\\
    Display impression type 4 & 160837\\
    Display impression type 5 & 355639\\
    Display click type 1 & 1700\\
    Display click type 2 & 161\\
    Display click type 3 & 1287\\
    Display click type 4 & 240\\
    Display click type 5 & 84\\
    Email sent type 1 & 2324088\\
    Email sent type 2 & 126088\\
    Email sent type 3 & 864342\\
    Email sent type 4 & 251481\\
    Email open type 1 & 1907397\\
    Email open type 2 & 65680\\
    Email open type 3 & 814296\\
    Email open type 4 & 180003\\
    Email click type 1 & 123190\\
    Email click type 2 & 2137\\
    Email click type 3 & 53556\\
    Email click type 4 & 8810\\
    Paid search click type 1 & 36123 \\
    Paid search click type 2 & 11781 \\
    Paid search click type 3 & 19139 \\
    Paid search click type 4 & 3240 \\

    \hline
\end{tabular}\label{table:touchpoints}

\begin{tablenotes}
\small
\item 
\end{tablenotes}
\end{table}

The data can be summarized by the input/output pairs as:
\begin{align}\label{eq:data}
\left\{\mathbf{x}^{(i)}\left[t^{(i)}-T,t^{(i)}\right],y^{(i)}\right\}_{i=1}^{N(T)}
\end{align}
where $N(T)$ is the total number of users and $\mathbf{x}\in\mathbb{R}^{31}$ is a 31-dimensional vector of touchpoint counts over the time period $[t(i)-T,t(i)]$ with $T$ being the length of the lookback period. We are interested in the lookback length, because of the complexity and costs of managing high-dimension, large-volume touchpoint data over long periods. We aim to assess the trade-offs associated with accepting a degree of inaccuracy by reducing the lookback windows for predictions. We will explore the potential impact on prediction accuracy when the firm opts to use data from a shorter lookback window, thereby providing valuable insights for managers about data use and operational efficiency.

If the user is a buyer, $t(i)$ is the time of purchase, and touchpoints after the first purchase are not included in the analysis. If the user is not a buyer, touchpoint counts from a randomly selected time period of length $T$ are used. A consumer is not included if they are not marketed in the chosen lookback window.
\section{Model}\label{sec:model}

Given the data \eqref{eq:data}, the goal is to learn a binary classifier, i.e., a function $C:\mathbb{R}^{31}\rightarrow\{0,1\}$ that accurately maps buyers to 1 and non-buyers to 0. To build such a function, this study leverages neural networks, which are parameterized functions that can be trained to approximate arbitrary nonlinear behavior. 

\subsection{Model Configuration}\label{subsec:config}

To begin, let $\mathbf{N}:\mathbb{R}^{31}\rightarrow(0,1)$ be a fully-connected neural network with $N_{l}$ hidden layers and $N_{n}$ nodes per hidden layer. We have conducted an extensive exploration of various configurations within our neural network model, including different loss functions, activation functions, and network architectures. 
Table \ref{table:configs} shows a \emph{summary} comparison of various configurations, although does not represent all tested configurations. Specifically, for loss functions, we examined three variations: binary cross entropy (BCE), binary focal cross entropy (BFCE) weighing ``hard to classify" training data more, and binary focal cross entropy weighing classes evenly despite imbalanced data. In terms of activation functions, we experimented with the sigmoid, rectified linear unit (ReLU), and hyperbolic tangent (tanh) functions. Regarding the network architecture, we evaluated models with 1 to 10 hidden layers containing up to 1000 nodes each. For the sake of brevity, Table \ref{table:configs} shows the results of a more local search around $N_l=3$ and $N_n=10$, which ultimately yielded the highest balanced accuracy. Moreover, our exploration extended to include variations in thresholding approaches and data-cleaning strategies. In terms of choosing an appropriate threshold, we tried maximizing the balanced accuracy (mean of the true positive and true negative rates) as well as maximizing the geometric mean of the true positive and true negative rates. For data cleaning, we considered two approaches: utilizing the entire dataset as is, or excluding the top and bottom 10\% of users based on cumulative touchpoint counts.

\begin{table}[ht]
\centering
\begin{threeparttable}
\caption{Neural Network Configuration Comparison}
\begin{tabular}{ l c c c c c c }
    \hline
Data & Loss & $N_l$ & $N_n$ & Params. & Activation & Balanced Accuracy \\
\hline
raw & BCE & 1 & 10 & 331 & sigmoid & 0.7702725 \\
raw & BCE & 2 & 10 & 441 & sigmoid & 0.7691675 \\
raw & BCE & 3 & 10 & 551 & sigmoid & \textbf{0.7760830} \\
raw & BCE & 4 & 10 & 661 & sigmoid & 0.7650520 \\
raw & BCE & 5 & 10 & 771 & sigmoid & 0.7589785 \\
\hline
raw & BCE & 3 & 5 & 226 & sigmoid & 0.7662415 \\
raw & BCE & 3 & 8 & 409 & sigmoid & 0.7707465 \\
raw & BCE & 3 & 12 & 709 & sigmoid & 0.7658305 \\
raw & BCE & 3 & 15 & 976 & sigmoid & 0.7629365 \\
\hline
raw & BFCE & 3 & 10 & 551 & sigmoid & 0.7650835 \\
raw &  BFCE w/ bal. & 3 & 10 & 551 & sigmoid & 0.7659045 \\
raw & BCE & 3 & 10 & 551 & ReLU & 0.5338935 \\
raw & BCE & 3 & 10 & 551 & tanh & 0.7613255 \\
cleaned & BCE & 3 & 10 & 551 & sigmoid & 0.7566020 \\
\hline
\end{tabular}
\label{table:configs}
     \end{threeparttable}
\end{table}

After exhaustively comparing the combinations of all these configurations, $N_l=3$ hidden layers with $N_n=10$ nodes each was chosen as the best-performing model configuration. This model utilizes binary cross entropy as the loss function and employs the sigmoid activation function. For thresholding, maximizing the sum of true positive and true negative rates proved to be more effective. Additionally, excluding the top and bottom 10\% of users based on cumulative touchpoint counts does not further improved the model's performance and therefore the raw data is used.

Specifically, the network function is
\begin{align}
    \mathbf{N}(\mathbf{x};\Theta) = \sigma\left(W_4\sigma\left(W_3\sigma\left(W_2\sigma\left(W_1\mathbf{x}+\mathbf{b}_1\right)+\mathbf{b}_2\right)+\mathbf{b}_3\right)+\mathbf{b}_4\right),
\end{align}
where
\begin{itemize}
    \item $\mathbf{x}\in\mathbb{R}^{31}$ is the input;
    \item $\mathbf{N}(\mathbf{x};\Theta)\in(0,1)$ is the output;
    \item $W_1\in\mathbb{R}^{31\times 10}$, $W_2,W_3\in\mathbb{R}^{10\times 10}$, and $W_4\in\mathbb{R}^{10\times 1}$ are the weight matrices;
    \item $\mathbf{b}_1,\mathbf{b}_2,\mathbf{b}_3\in\mathbb{R}^{10}$ and $\mathbf{b}_4\in\mathbb{R}$ are the bias terms;
    \item and $\sigma(t) = \frac{1}{1+e^{-t}}$ is the sigmoid activation function.
\end{itemize}
Collectively, $\Theta\in\mathbb{R}^{551}$ represents all of the 551 trainable parameters, i.e., the elements of weight matrices and bias terms. A training procedure, detailed in Section \ref{subsec:training}, chooses optimal parameters $\Theta^*$.

The randomness in neural network training, which arises from random parameter initialization and stochastic optimization, can also be harnessed to build a more accurate classifier. By training an ensemble of $K$ models, where in this case $K=10$, with each model beginning with a different random seed, and averaging their results, an ensemble model is:
\begin{align}
    \mathbf{N}_{ens}(\mathbf{x}) = \frac1K\sum_{k=1}^K \mathbf{N}_k(\mathbf{x}; \Theta^*_k),
\end{align}
where $\mathbf{N}_k$ and $\Theta_k^*$ correspond to the network function and optimal parameters of each of the $K$ constituent models.

\begin{figure}[htbp]
	\begin{center}
 \caption{Diagram of the Proposed Binary Classification Model}
        \includegraphics[width=1.02\textwidth]{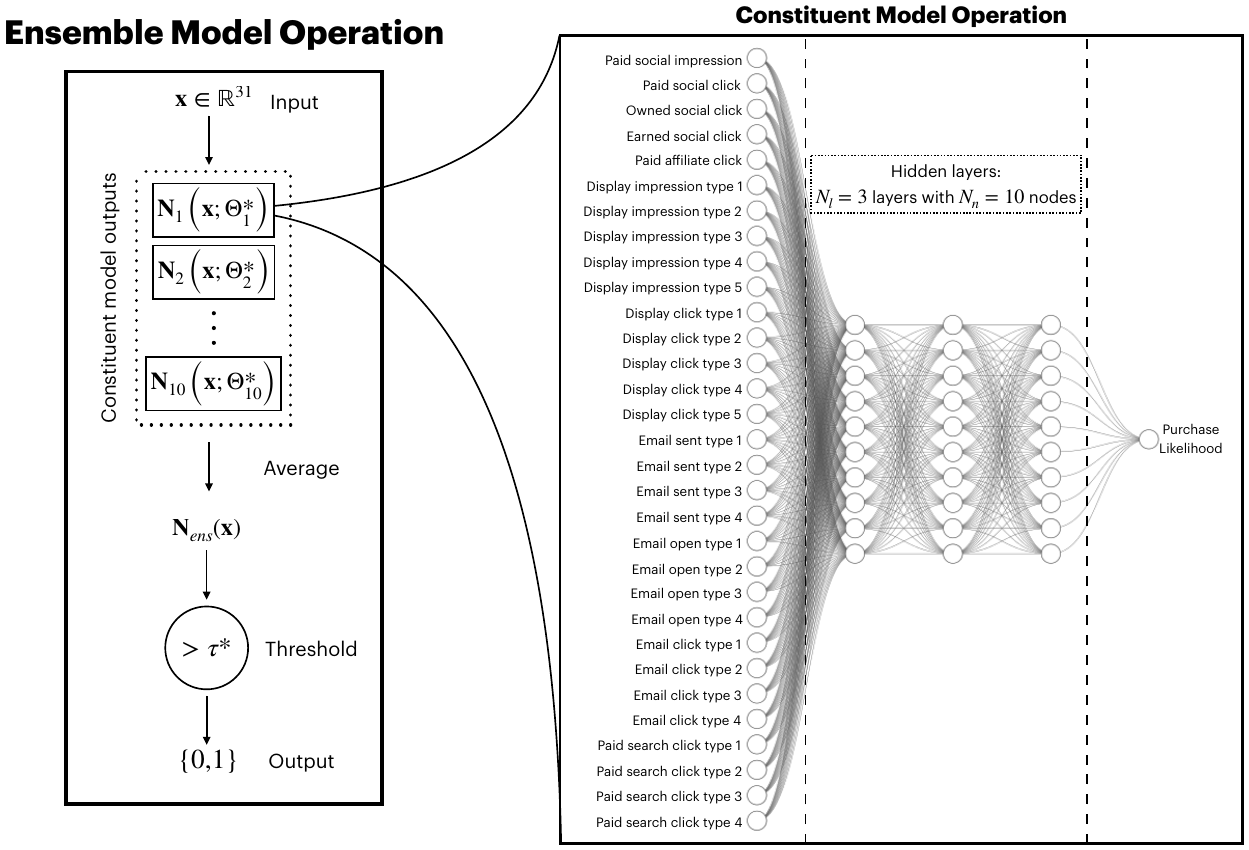}
		\label{fig:diagram}
	\end{center}
\end{figure}

Additionally, a threshold $\tau^*\in(0,1)$ must be chosen in order to assign a binary 0 or 1 to the network output. Therefore, the function
\begin{align}\label{eq:bc}
    C(\mathbf{x}) = \begin{cases} 1 & \mathbf{N}_{ens}(\mathbf{x})>\tau^* \\ 0 & \mathbf{N}_{ens}(\mathbf{x}) \le \tau^* \end{cases}
\end{align}
represents the binary classifier. Figure \ref{fig:diagram} gives a visual representation of the full model operation.

\subsection{Training}\label{subsec:training}
In order to cross-validate the model, the data \eqref{eq:data} is separated randomly into 80\% training, 10\% validation, and 10\% testing sets. Each constituent network is trained, i.e., parameters $\Theta$ are chosen to match the input/output behavior exhibited by the data, by minimizing the binary cross-entropy loss function
\begin{align}
    \arg\min_\Theta \left\{ -\frac1N \sum_{i=1}^N \left[y^{(i)} \log \left(\mathbf{N}\left(\mathbf{x}^{(i)};\Theta\right)\right) + \left(1-y^{(i)}\right)\log\left(1-\mathbf{N}\left(\mathbf{x}^{(i)};\Theta\right)\right) \right]\right\}
\end{align}
over the training dataset. The Adam optimizer, \citep{kingma2014adam}, is used to reduce the loss over $10,000$ epochs with a constant learning rate of $10^{-3}$. Throughout the optimization, the parameters $\Theta$ change. In order to ensure the models generalize well to new data, the validation dataset, which is not used to minimize the loss, is used to choose one of these parameterizations. A parameterization $\Theta^*$ is chosen for each constituent network by maximizing the AUROC over the validation data.

Once the networks have been trained, a threshold $\tau$ is chosen from the ensemble ROC curve. Since the classes are imbalanced, the threshold is chosen as the maximizer of the \emph{balanced} accuracy
\begin{align}\label{eq:threshold}
    \tau^* = \arg\max_{\tau\in(0,1)}\left\{ \frac{TPR(\tau)+TNR(\tau)}{2} \right\}
\end{align}
where $TPR$ and $TNR$ are the true positive and true negative rates on the validation data at $\tau$, respectively.\footnote{While the balanced accuracy is the arithmetic mean of the $TPR$ and $TNR$, choosing the threshold via the geometric mean was also tested for all configurations in Table \ref{table:configs} and did not yield superior performance.} At this point, the binary classifier \eqref{eq:bc} is fully trained and the test dataset can be used to compare results.
\section{Predictive Performance}\label{sec:results}

To demonstrate the enhanced predictive performance of our proposed neural network model, it undergoes comparative analysis against several established binary classifiers: logistic regression, naive Bayes, and $k$-nearest neighbors. Additionally, it was tested against state-of-the-art ensemble tree-based classifiers such as bagging, random forest, AdaBoost, and gradient boosting. For ensemble methods, 10 models were used to present a fair comparison against our 10-model ensemble. Details on these standard methods can be found in  \cite{hastie2009elements}. Implementation of competing models was performed according to \cite{scikit-learn}. Identical training, validation, and test data are used.

\begin{figure}[htbp]
	\begin{center}
 \caption{ROC Curves for All Models Using a 40-month Lookback Window}
        \includegraphics[width=.9\textwidth]{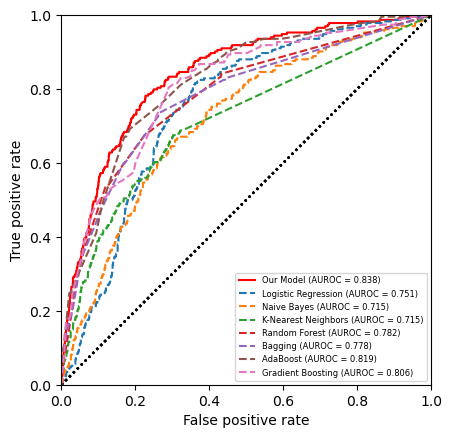}

		\label{fig:ROC}
	\end{center}
 \caption*{{\small Note: AUROC stands for the area under the receiver operating characteristics (ROC) curve.}}
\end{figure}

Figure \ref{fig:ROC} displays the ROC curves for our proposed model and the aforementioned benchmark models using 40 months of data. With an AUROC of 0.838, our model surpasses all other models in accurately classifying positive instances over negative ones. Models such as AdaBoost, gradient boosting, and random forest show commendable performance, though lower than our model. Other models like bagging, logistic regression, k-nearest neighbors, and naive Bayes exhibit progressively lower AUROC values, indicating less predictive accuracy.

Table \ref{table:metrics_alltime} 
further presents the accuracy metrics for our proposed model versus the benchmark models. Alongside the AUROC values depicted in Figure \ref{fig:ROC}, this table also reports the TPR, TNR, and Balanced Accuracy, with the latter being the mean of TPR and TNR as delineated in \eqref{eq:threshold}. We believe marketing managers should evaluate both TPR and TNR when gauging campaign effectiveness. While the benefits of a high TPR, which ensures that those most likely to convert are accurately targeted, are widely recognized, the significance of TNR cannot be overstated. Given the prevalence of non-buyers in most CRM databases and the even scarcer presence of buyers in short-term windows, a high TNR becomes crucial. It prevents the wasteful expenditure of marketing resources on individuals unlikely to convert, and thus helps managers maintain a cost-effective marketing strategy, mitigate customer irritation from irrelevant outreach, and uphold a favorable brand image by avoiding excessive communication. Achieving a balance between TPR and TNR offers a comprehensive efficiency measure, enabling marketing managers to refine their strategies towards both precision in targeting and budget efficiency.

Our model shows balanced and robust rates of both TPR and TNR, at 0.782051 and 0.770115, respectively, leading to the highest Balanced Accuracy of 0.776083. While the Gradient Boosting model exhibits a strong TPR, it does not achieve a comparable TNR, leading to an overall lower performance than the proposed model, as reflected in both the AUROC and Balanced Accuracy metrics. Random Forest, while having a slightly lower AUROC at 0.800007, shows the best performance in correctly identifying negatives with a TNR of 0.869184 but falls short in TPR with 0.538462. Overall, these metrics emphasize our model's superior and balanced classification performance.

\begin{table}[ht]
\centering
\caption{Accuracy Metrics for All Models Using a 40-month Lookback Window}
\begin{tabular}{ l c c c c }
    \hline
& AUROC & TPR & TNR & Balanced Accuracy \\
\hline
Our Model & \textbf{0.837972} & 0.782051 & 0.770115 & \textbf{0.776083} \\
Logistic Regression & 0.751428 & 0.743590 & 0.687466 & 0.715528 \\
Naive Bayes & 0.714832 & 0.692308 & 0.637110 & 0.664709 \\
K-Nearest Neighbors & 0.715041 & 0.688034 & 0.674330 & 0.681182 \\
Random Forest & 0.800007 & 0.538462 & \textbf{0.869184} & 0.703823 \\
Bagging & 0.777554 & 0.598291 & 0.829228 & 0.713759 \\
AdaBoost & 0.819020 & 0.760684 & 0.729611 & 0.745148 \\
Gradient Boosting & 0.806103 & \textbf{0.811966} & 0.694581 & 0.753274 \\
\hline
\end{tabular}
\label{table:metrics_alltime}
   \begin{tablenotes}
    \small
  \item[]\hspace{1em}Note: Bold indicates the highest value in each column.
    \end{tablenotes}
\end{table}
\section{Discussion}
\subsection{Interpreting the Importance of Touchpoints}
\begin{figure}[htbp]
	\begin{center}
 \caption{Beeswarm Plot of Shapley Values}
        \includegraphics[width=.74\textwidth]{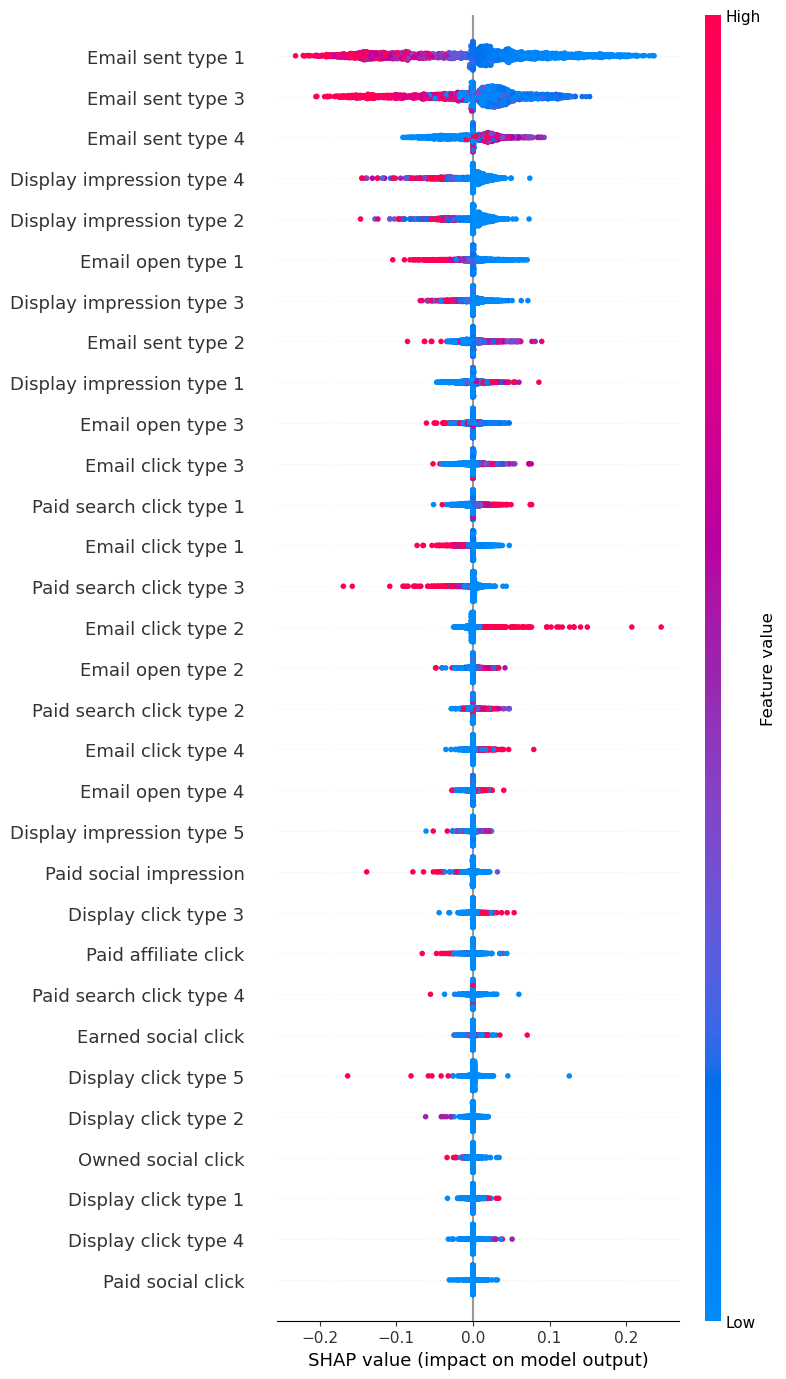}
		\label{fig:shap}
	\end{center}
\end{figure}

 The Shapley value was originally developed by \cite{shapley1953value} to allocate a fair payoff within cooperative game theory, based on each player's contribution to the total payoff. In our context, we calculate the Shapley values of each touchpoint type to quantify their effect on the outcome, namely consumer purchases, for each consumer in the test dataset. The beeswarm plot of Shapley values depicted in Figure \ref{fig:shap} provides a visualization of the influence that various marketing touchpoints have on the likelihood of purchase. This analysis is conducted using the SHAP library in Python, following the implementation set out by \citep{NIPS2017_7062}, which formalizes the Shapley values in a machine learning context.

In Figure \ref{fig:shap}, each consumer in the test dataset is represented as a dot in each row, with each row representing a different touchpoint type. The position on the x-axis of the dot is determined by the Shapley value of that touchpoint type. Dots to the right of the zero indicate a higher likelihood of the model predicting a purchase, while dots to the left suggest a lower likelihood. The color of each dot reflects the touchpoint count for that user: red for higher and blue for lower. Dots cluster along each row to show density. Along the y-axis, touchpoint types are ordered by their overall importance, i.e. the sum of absolute Shapley values over all consumers, with the most important at the top. For example, ``Email sent type 1'' is shown to be the most important touchpoint type on average. Interestingly, within this row, a predominance of blue dots on the right for this touchpoint suggests fewer touchpoints of “Email sent type 1” are associated with a higher purchase likelihood, implying possible diminishing returns or adverse effects beyond a certain frequency.

The details in Figure \ref{fig:shap} offer valuable insights for the firm. For example, regarding the impact of ``Email sent types'' on purchase likelihood, Types 1 and 3 are found to be negatively associated with buying decisions, whereas Types 2 and 4 have a positive correlation with purchases. However, if a consumer clicks on the email, ``Email click types" 2 through 4 all enhance conversion rates, while Type 1 exhibits mostly negative associations.

Within “Display impressions,” Types 1 through 4 exhibit relatively higher importance among the touchpoint types in predicting purchases, but not Type 5. Meanwhile, Type 1 shows a positive relationship with purchasing likelihood, unlike the negative or mixed correlations of the other types. Moreover, all types of “Display clicks” have a low overall impact on purchases. Types 1, 3, and 4 are mostly positive, while Types 2 and 5 mainly relate negatively to purchases.

“Paid search click types” 1 through 3 are of moderate importance in predicting purchases, whereas Type 4 is of lower importance. Types 1 and 2 show a positive association, while Types 3 and 4 exhibit negative associations. 

Our Shapley value analysis provides nuanced insights into the influence of various touchpoints on consumer purchases. It not only delineates the relative importance of each touchpoint type, but also illustrates the distribution within types, offering a comprehensive view of their impact on purchase decisions. The analysis reveals that certain types of touchpoints can either positively or negatively affect purchases, which is pivotal for firms to learn before optimizing their marketing spending. Specifically, it appears the focal firm may be overutilizing some marketing efforts, while their click-based interactions show more positive impact than impression-based touchpoints.

\subsection{Shorter Lookback Windows}
This section further explores the robustness of our proposed model by examining its predictive capability with shorter lookback windows. Collecting, maintaining, and analyzing high-dimensional touchpoint data entails significant costs for firms. While our main analysis leverages an extensive dataset covering 40 months, we aim to delve into the model’s effectiveness when utilizing shorter historical data periods. This investigation responds to the needs of many firms that may lack the resources to collect, maintain, and analyze data over such extended durations. By evaluating the model’s performance with more constrained data sets, we seek to demonstrate its applicability and value to a broader range of business contexts.

To address this, we reduce our lookback window from the extensive 40-month period to more condensed intervals such as 12 months, 3 months, or just 1 month, and assess whether our model remains superior under constrained data conditions. The outcomes confirm the robust accuracy of our model. However, due to limited space, we only present the accuracy metrics for 1-month lookback window in Tables \ref{table:metrics_month}, where our model presents the best AUROC of 0.675199 and maintains the highest Balanced Accuracy at 0.630247 among all comparative models. 

It's compelling to observe that as we transition from a 40-month to a 1-month period, the decline in AUROC is a modest 0.162773 (or 19.4\%), and Balanced Accuracy sees a similarly modest reduction of 0.145836 (or 18.8\%). This moderate decrease in accuracy is in sharp contrast against the substantial benefits of reducing data storage needs by a staggering 97.5\% and significantly decreasing the complexity of data analysis, a trade-off that many firms may find not only acceptable but highly advantageous. 

Our model's consistent accuracy with shorter data periods confirms its robustness, making it a valuable tool for companies aiming for efficient predictive modeling. It is particularly useful for new or dynamic businesses where long historical data may be scarce or quickly outdated. Firms can leverage this model to make quick, informed decisions and stay agile with minimal data management overhead.

\begin{table}[ht]
\centering
\caption{Accuracy Metrics for All Models Using a 1-month Lookback Window}
\begin{tabular}{ l c c c c }
    \hline
& AUROC & TPR & TNR & Balanced Accuracy \\
\hline
Our Model & \textbf{0.675199} & 0.530702 & 0.729792 & \textbf{0.630247} \\
Logistic Regression & 0.666103 & \textbf{0.592105} & 0.640878 & 0.616491 \\
Naive Bayes & 0.652258 & 0.447368 & 0.762125 & 0.604747 \\
K-Nearest Neighbors & 0.578704 & 0.403509 & 0.725751 & 0.564630 \\
Random Forest & 0.572877 & 0.500000 & 0.577945 & 0.538972 \\
Bagging & 0.621032 & 0.500000 & 0.685335 & 0.592667 \\
AdaBoost & 0.668625 & 0.390351 & 0.854503 & 0.622427 \\
Gradient Boosting & 0.627334 & 0.324561 & \textbf{0.890300} & 0.607431 \\
\hline
\end{tabular}
\label{table:metrics_month}
   \begin{tablenotes}
    \small
  \item[]\hspace{1.1em}Note: Bold indicates the highest value in each column.
    \end{tablenotes}
\end{table}

\section{Implications and Future Research}

In the ever-evolving landscape of digital marketing, firms navigate through extensive consumer touchpoint data, essential for interpreting consumers’ purchase decisions. However, marketing managers face the demanding challenge of analyzing large volume and high-dimensional data spanning various channels to influence consumer behavior effectively. Conventional statistical models frequently struggle in this context, burdened by the voluminous scale and dimension of data, which can render them impracticable for nuanced analysis.

Our model employs automated learning through neural networks to uncover hidden patterns within consumer touchpoint data. This approach grants marketing managers a quicker and more precise insight into the customer journey, pinpointing those prospects who are most and least likely to convert. By leveraging this model, firms can better synchronize their marketing strategies with consumer interactions, thus boosting the effectiveness and efficiency of their marketing campaigns.

This research illustrates the superior accuracy of neural networks in analyzing the consumer purchase journey, surpassing traditional models in predicting binary purchase outcomes. We evaluate our neural network model against established binary classifiers, including logistic regression, naive Bayes, and k-nearest neighbors, as well as sophisticated ensemble tree-based classifiers such as bagging, random forest, AdaBoost, and gradient boosting. Our neural network framework outperforms all these alternative methods, achieving superior performance as evidenced by the highest AUROC as a measure of the model’s ability to distinguish between classes and, meanwhile, the highest balanced accuracy, which considers both the TPR and TNR.

We utilize Shapley values to evaluate the effect of the 31 touchpoint types on consumer purchases, both ranking touchpoint types by their importance and exploring their distribution, offering granular insights into their role in purchasing decisions. The proposed approach is particularly beneficial for firms aiming to optimize their marketing strategies by understanding which touchpoint types drive consumer purchases. Firms in highly competitive sectors, such as e-commerce, technology, and services, can leverage these insights to fine-tune their customer engagement tactics, ensuring resources are directed towards the most effective channels for enhancing consumer interaction and boosting sales.

Considering the challenges of high-dimensional data, our model shows strong adaptability and resilience when performed with only one month of historical data, balancing data management ease with predictive precision, thereby aiding firms in strategic decision-making. The proposed model is valuable for both large enterprises with vast amounts of consumer data and smaller companies seeking to maximize limited marketing resources and data access. 

We recognize that the architecture of our proposed neural networks is tailored to our specific dataset. Although ours favored a simpler, shallow network, different scenarios might benefit from deeper networks. Our exploration of the model with shorter lookback windows, from 40 months down to just 1 month, indicates its robust predictive power even with limited historical data, suggesting that extensive historical data may not be crucial for accurate purchase predictions in our context. Generally speaking, feedforward neural networks are often preferred over long short-term memory (LSTM)  networks, a variant of recurrent neural networks, due to the former’s ability to handle inputs independently and more efficient parallel computation capabilities. Feedforward networks also tend to have fewer parameters compared to the gate-laden structure of LSTMs, resulting in less computational intensity and shorter training periods. In future applications, researchers may examine the importance of sequence in that context and determine whether the increased computational burden of LSTMs is justified. Another potential avenue for future research is integrating unstructured data, such as text, images, and videos, into the model to provide richer insights into consumer purchase journeys.

\newpage

\section*{Funding and Competing Interests}
All authors certify that they have no affiliations with or involvement in any organization or entity with any financial interest or non-financial interest in the subject matter or materials discussed in this manuscript. The authors have no funding to report.

\newpage
\bibliographystyle{ormsv080}  
\bibliography{refs.bib}


\end{document}